%% file: main.tex
\documentclass[letterpaper,10pt]{article} 

\usepackage{opticameet3} 

\newcommand\authormark[1]{\textsuperscript{#1}}

\usepackage{amsmath,amssymb}
\usepackage[colorlinks=true,bookmarks=false,citecolor=blue,urlcolor=blue]{hyperref} 
\usepackage{graphicx}
\usepackage{tikz}
\usepackage{pgfplots}
\usepackage{mathtools}
\usepgfplotslibrary{colorbrewer}
\usepgfplotslibrary{groupplots}
\usetikzlibrary{shapes.geometric}
\usepackage{pgfplotstable}
\pgfplotsset{compat=1.17}
\usetikzlibrary{backgrounds}
\usepackage{wrapfig}  
\usepackage{subfigure}
\usepackage{amsfonts}
\usepackage{multirow}
\usetikzlibrary{arrows}
\usetikzlibrary{decorations.markings}
\usetikzlibrary{matrix}
\usepackage{xcolor}
\usepackage{colortbl}
\usepackage{threeparttable}
\usepackage{environ}
\usepackage{multicol}
\newcommand{\ASE}{\ensuremath{\text{ASE}}}

\newcommand{\SNR}{\ensuremath{\mathrm{SNR}}}

\begin{document}

\title{Throughput Maximisation in Ultra-wideband Hybrid-amplified Links}


\author{Henrique~Buglia,\authormark{1} Eric~Sillekens,\authormark{1}  Lidia~Galdino,\authormark{2} Robert~Killey\authormark{1} and Polina~Bayvel\authormark{1}}

\address{\authormark{1} Optical Networks Group, Dept. of Electronic \& Electrical Engineering, University College London, U.K. \\
\authormark{2} Corning Optical Communications, Ewloe, U.K}

\email{\authormark{}henrique.buglia.20@ucl.ac.uk} 

\vspace{-0.4cm}
\begin{abstract}
A semi-analytical, real-time nonlinear-interference model including ASE noise in hybrid-amplified links is introduced. Combined with particle-swarm optimisation, the capacity of a hybrid-amplified 10.5-THz 117x57-km link was maximised, increasing throughput by 12\% versus an EDFAs-only configuration.
\end{abstract}

\section{Introduction}\vspace{-0.2cm}

To enable the exponential growth of data transmission required by new internet services, ultra-wideband (UWB) transmission technologies have been widely explored in optical fibre communication systems. Among these technologies, the use of hybrid-amplification schemes was almost universally applied in all the recent field trials achieving record throughputs over single-mode fibre (SMF), as shown in~\cite{BugliaJOCN}. The increased throughput with hybrid amplification is due to the lower noise figure achievable with distributed Raman amplification in comparison to that possible using Erbium-doped fibre amplifiers (EDFA) alone. Distributed amplification also allows a reduction in the launch power and correspondingly - in the nonlinear interference (NLI) noise. To optimise this type of system, pump optimisation is required to find the best wavelength and power allocation to maximise the system throughput. To achieve this, real-time estimation of the nonlinear interference (NLI) noise in Raman-amplified links is required, enabled by recently-developed semi-analytical expressions of the Gaussian noise (GN) model in the presence of Raman amplification (RA)~\cite{RamanJLT,ECOC_Raman,bugliaOFC, torino}. Together with this, the amplified spontaneous emission (ASE) noise generated by Raman amplifiers needs to be taken into account in this model to obtain a complete estimation of the total received signal-to-noise ratio (SNR). 

In this paper, for the first time, we have included the ASE noise generated by Raman amplifiers in the model~\cite{RamanJLT,ECOC_Raman,bugliaOFC}. This new step allowed the design of a capacity-achieving backward-pumped hybrid amplifier.  Using particle swarm optimisation (PSO) algorithm to optimise the total launch power, pump powers and pump wavelengths resulted in a maximum system throughput within a 10.5~THz transmission bandwidth, corresponding to the utilisation of the C- and L-band, where backward Raman pumps are placed in the S-band. The results are finally compared with the same system optimised to operate with EDFAs alone. This work represents the first complete (ASE + NLI) semi-analytical model which allows the calculation of the received SNR \& throughput in real time for any UWB Raman-amplified link setup.

\vspace{-0.15cm}
\section{The GN Model in the Presence of RA and the Raman ASE Noise Generation}
\vspace{-0.2cm}

For an ideal transceiver, the SNR for the $i$-th channel at the end of the span, after amplification, can be estimated as $\text{SNR}_{i}^{-1} \approx \text{SNR}_{\text{NLI},i}^{-1} + \text{SNR}_{\text{ASE},i}^{-1}$, where 
$\text{SNR}_{\text{NLI},i}$, $\text{SNR}_{\text{ASE},i}$ originate from fibre nonlinearity and amplifier noise, respectively. In this work, ideal amplification was assumed, so that the transmitted power is fully recovered at the receiver. For Raman amplifiers, the evolution of the channel of interest (COI) power along the fibre distance ($P_i(z)$) is coupled with the ASE noise, generated by the remaining channels and the pumps, and is written as
\begin{equation}
\vspace*{-.20cm}
\small
\begin{aligned}
 \pm  & \frac{\partial P_i}{\partial z} =  - \sum_{k=i+1}^{\text{N}_{\text{ch}}} \frac{f_k}{f_i} g(|\Delta f|) (P_k + P_{\ASE,k}) P_i - \sum_{p:f_i>f_p} \frac{f_p}{f_i} g(|\Delta f|) (P_p + P_{ASE,p}) P_i + \\ &\sum_{k=1}^{i-1} g(|\Delta f|) (P_k + P_{\ASE,k}) P_i + \sum_{p:f_i<f_p} g(|\Delta f|) (P_p + P_{\ASE,p}) P_i - \alpha_i P_i,
\end{aligned}
\label{eq:diff_Raman}
\end{equation}
where, $P_i$, $f_i$ are the power and frequency of the COI, $P_k$, $f_k$ are the power and frequency of the remaining WDM channels, $P_p$, $f_p$ are the power and the frequency of the pumps, $g_r(|\Delta f|)$ is the polarisation averaged, normalized (by the effective core area $A_\text{eff}$) Raman gain spectrum for a frequency separation $|\Delta f| = |f_i - f_j|$, $j=k,p$ and $\alpha_i$ is the frequency-dependent attenuation coefficient. The symbol $\pm$ represents the pump under consideration, i.e., $+$ for forward pump and $-$ for backward pump configurations. $P_{\ASE,i}$, $P_{\ASE,k}$ and $P_{\ASE,p}$ are the ASE noise respectively in the COI, channel $k$ and pump $p$. The $\text{SNR}_{\text{NLI},i}$ is calculated in closed form using the model in~\cite{RamanJLT,ECOC_Raman}, which is derived from the integral ISRS~GN~model~\cite{Semrau_integral}, by using a suitable semi-analytical expression to express the normalised power profile evolution along the fibre distance, given by $\rho(z,f_i) = P_i(z)/P_i(0)$ (see Appendix~A of ~\cite{RamanJLT}), where $P_i(z)$ is obtained from Eq~\eqref{eq:diff_Raman}. The ASE noise generated in channel $i$ ($P_{\ASE,i}$) is obtained as~\cite{RamanEquations}
\begin{equation}
\vspace*{-.1cm}
\begin{aligned}
  &\frac{\partial P_{ASE,i}}{\partial z} = - \sum_{k= i + 1}^{\text{N}_{\text{ch}}}\frac{f_k}{f_i} g(|\Delta f|) (P_k + P_{ASE,k}) (P_{ASE,i} + 2h\kappa B_if_i) -\sum_{p: f_i > f_p}\frac{f_p}{f_i} g(|\Delta f|) (P_p + P_{ASE,p}) (P_{ASE,i} + 2h\kappa B_if_i) + \\
  & + \sum_{k=1}^{i-1} g(|\Delta f|) (P_k+ P_{ASE,k}) (P_{ASE,i} + 2h\kappa B_if_i) + \sum_{p: f_i < f_p} g(|\Delta f|) (P_p+ P_{ASE,p}) (P_{ASE,i} + 2h\kappa B_if_i) - \alpha_i P_{ASE,i},
\end{aligned}%
\label{eq:diff_Raman_ase}
\end{equation}
with $\kappa = 1 + \eta = 1/(1 - exp(-h\Delta  f/k_B/T))$, where $\eta$ is the phonon occupancy factor, $h$ is the Planck constant, $T$
is the temperature of the system and $k_B$ is Boltzmann’s constant. Note that Eq.~\eqref{eq:diff_Raman} should be written for each channel $k$ and pump $p$ by replacing $i=p,k$, whereas Eq.~\eqref{eq:diff_Raman_ase} is written for each channel $k$ by replacing $i=k$. Together, they represent a set of $2N_{ch} + N_{p}$ coupled differential equations describing the signal evolution and the ASE noise generation in Raman-amplified links. This equation is solved for each span, where the accumulated ASE noise at the end of each span is used as the initial condition for the following span. For hybrid-amplified links, the ASE generated by distributed Raman amplification, obtained from Eq.~\eqref{eq:diff_Raman_ase}, is amplified by the ideal EDFA gain ($G_i$) placed at the end of the fibre.
The total ASE noise is then given by $P_{\ASE,i}^{\prime \prime} = G_i P_{\ASE,i} + P_{\ASE,i}^\prime$, where $P_{\ASE,i}^\prime = 2(G_i-1)n_{sp} hf_iB_i$, with $n_{sp} \approx \text{NF}/2$ the spontaneous emission factor and $\text{NF}$ the EDFA noise figure. The per-channel SNR contribution from the total ASE noise is then calculated as $\text{SNR}_{\text{ASE},i} = P_i/P_{\text{ASE,i}}^{\prime \prime}$. 
\vspace{-0.2cm}
\section{Transmission Setup and System Optimisation}
\label{Transmission Setup and Throughput Maximisation}\vspace{-0.2cm}

\begin{figure}[t!]
\hspace*{-.8cm}
\vspace*{-.3cm}
    \input{Figures/attenuation_Gain}
\caption{(a) Fibre attenuation coefficient and Raman gain spectrum for the fibre used; and optimised pump wavelength allocation. (b) Optimised hybrid amplifier gain for backwards distributed Raman amplification stage (red) and ideal lumped (EDFA) stage (blue).}
\label{fig:attenuation_gain}
\end{figure}
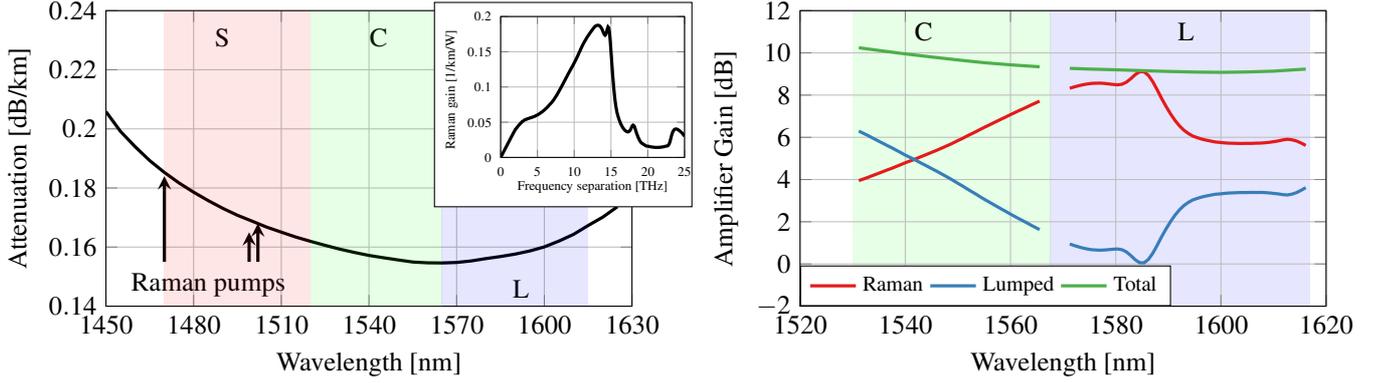

The transmission system is assumed to use a hybrid amplification technology, consisting of two stages: a backward-distributed Raman amplifier followed by an ideal EDFA. It was assumed to amplify a WDM signal with $N_{\text{ch}}$=105 Nyquist-spaced channels centred at 1571~nm, where each channel was modulated at the symbol rate of 100~GBd, with Gaussian symbols. This setup results in a total bandwidth of 10.5~THz, ranging from 1530~nm to 1615~nm, corresponding to the full utilisation of the C- and L- bands. 
Signal transmission was evaluated over 117~x~57~km (a total distance of 6669~km). The transmission optical fibre under consideration is assumed to have wavelength-dependent attenuation and the Raman gain profile compliant with Fig.~\ref{fig:attenuation_gain}\textcolor{red}{.a}, and an effective area of $150~\mu\text{m}^2$, resulting in a nonlinear coefficient $\gamma = 0.55\text{ W}^{-1} \text{km}^{-1}$. 
Dispersion parameters considered are $D = 21~\text{ps }\text{nm}^{-1}\text{km}^{-1}$, $S = 0.067~\text{ps }\text{nm}^{-2}\text{km}^{-1}$. The NFs of the EDFAs were assumed to be 5~dB and 6~dB, for the C- and L- bands, respectively. For the distributed Raman amplification, pumps were placed in the S-band, as shown in Fig.~\ref{fig:attenuation_gain}\textcolor{red}{.a}, and their wavelengths and powers, as well as the total transmitted launch power are optimised to maximise the system throughput.

The optimised hybrid amplifier was designed, based on a PSO algorithm, where a spectrally uniform launch power profile and 6 backward pumps, limited to $500~\text{mW}$ each and placed in the S-band (1470~nm - 1520~nm) are considered. The total launch power and the pump wavelengths and powers are optimised to maximise the cost function $C = \sum_{i = 1}^{N_{\text{ch}}} 2\cdot\log_2(1+\SNR_i)$, such that the total throughput is maximised. The PSO algorithm has 7 variables to be optimised (6 pumps + total launch power). The number of particles was chosen to be 50 with a maximum of 50 iterations selected as the stopping criterion. For the algorithm bounds, we let the total channel launch power vary between 15~dBm and 25~dBm, and the power of each pump at the end
of the fibre from 0~mW to 500~mW. The optimisation resulted in a total launch power of 20.4~dBm, corresponding to 0.49~dBm per channel, and 3 pumps with non-negligible power, with wavelengths 1470~nm, 1499~nm and 1502~nm, and powers of 433~mW, 107~mW and 113~mW, respectively as shown in Fig.~\ref{fig:attenuation_gain}\textcolor{red}{.a}. The optimised hybrid amplifier gain is shown in Fig.~\ref{fig:attenuation_gain}\textcolor{red}{.b} where the lumped (EDFA) gain is assumed to be ideal to completely recover the transmitted power. 

\vspace{-0.2cm}
\section{Throughput Maximisation Results}
\vspace{-0.2cm}

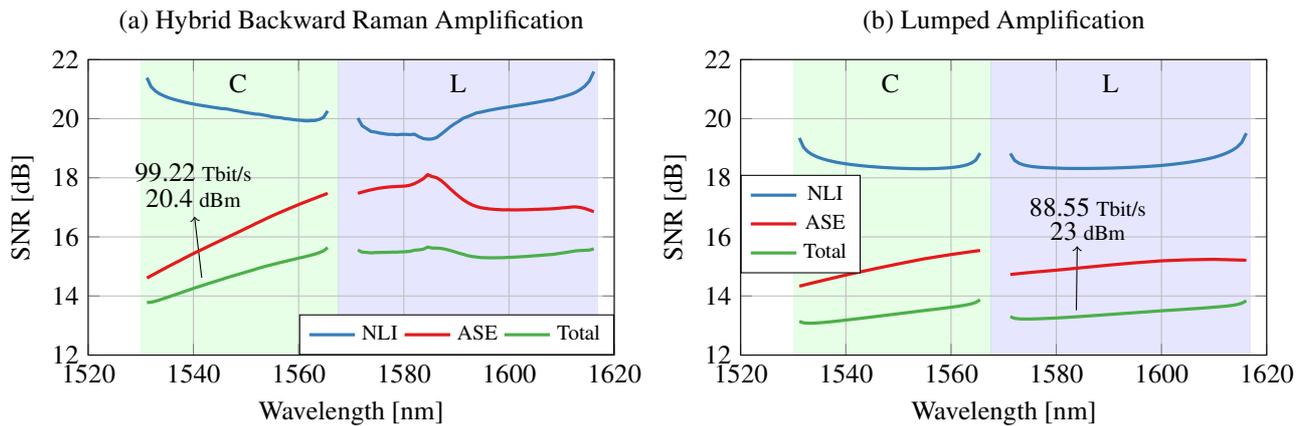
\begin{figure}[t!]
\hspace*{-.8cm}
\vspace*{-.3cm}
    \input{Figures/SNR}
\caption{Performance in terms of SNR for the system in Sec.~\ref{Transmission Setup and Throughput Maximisation} operating with an optimised (a) hybrid amplification and (b) full lumped amplification.}
\label{fig:SNR}
\end{figure}

Fig~\ref{fig:SNR}\textcolor{red}{.a} shows the transmission system performance of the capacity-achieving hybrid amplifier in terms of \text{SNR}. Because the majority of Raman gain occurs at the shorter wavelengths of the L band (see Fig.~\ref{fig:attenuation_gain}\textcolor{red}{.b}), the NLI noise is worst in this region, due to the increased power levels propagating through the fibre, which reduces the $\text{SNR}_{\text{NLI}}$. On the other hand, the better ASE performance of Raman amplifiers when compared to EDFA reduces the ASE noise in this same region increasing the $\text{SNR}_{\text{ASE}}$; this is because for those channels almost all the fibre loss is compensated with RA, whereas in the remaining part of the spectrum, greater EDFA gain is required.

To compare the benefits of using hybrid amplification schemes, Fig~\ref{fig:SNR}\textcolor{red}{.b} show the performance results in terms of SNR for the same transmission system described in Sec.~\ref{Transmission Setup and Throughput Maximisation} using a fully lumped amplification scheme, i.e, without any distributed Raman pumps injected in the transmission fibre. The results are calculated using the model in~\cite{daniel_closed} and the total launch power was also optimised to maximise the throughput, resulting in a value of 23~dBm, corresponding to 3.09~dBm per channel. The increased ASE noise values for this case when compared to the hybrid amplification case are a result of the lower performance of EDFAs in comparison to Raman amplifiers in terms of NF. In the case of NLI noise, despite pump powers also being injected into the transmission fibre for the hybrid amplification case, the increased channel launch powers optimising the system with lumped amplification make this case also worse in terms of NLI noise. Indeed, with optimised hybrid amplification, a total throughput of 99.22~Tbit/s was obtained with the total channel launch power of 20.4~dBm, while for the system with optimised lumped amplification, this value was 88.55~Tbit/s with 23~dBm total launch power. 
\vspace{-0.2cm}
\section{Conclusions}
\vspace{-0.2cm}
The first semi-analytical, real-time nonlinear-interference model capable of predicting the performance of hybrid-amplified (EDFA + Raman) systems was developed. The key new step was the inclusion of the Raman ASE noise in this model. Enabled by this, the throughput of a 10.5-THz 6669-km link was maximised by using a particle-swarm optimisation (PSO) algorithm. A total throughput of 99.22~Tbit/s was achieved, corresponding to a 12~\% increase when compared to the same system operating with EDFAs only. 
\vspace{0.2cm}
\\
\footnotesize{
\textbf{Financial support} under the EPSRC Programme Grant TRANSNET (EP/R035342/1) and EWOC (EP/W015714/1),  EPSRC studentship (EP/T517793/1), and the Microsoft 'Optics for the Cloud' Alliance are gratefully acknowledged.} 
\vspace*{-0.2cm}
\bibliographystyle{opticajnl}
\bibliography{sample}
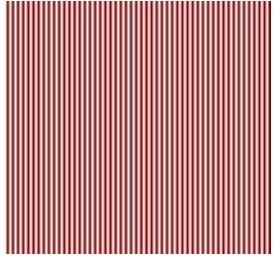
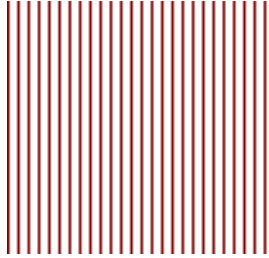
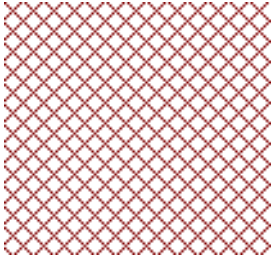
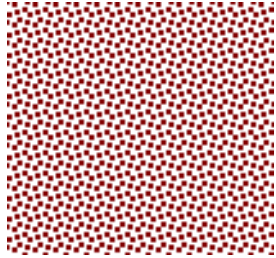
\end{document}

%% file: Figures/attenuation_Gain.tex
\begin{tikzpicture}[baseline]
\begin{axis}[
legend cell align=left,
title style={at={(axis cs:1540,0.239)}},
legend style={font=\footnotesize, at={(rel axis cs:0,1)}, anchor=north west},
width=8.5cm, height = 5.5cm,
xlabel={Wavelength [nm]},
ylabel={Attenuation [dB/km]},
grid=both,
xtick={1450,1480,1510,1540,1570,1600,1630},
ymax=0.24,ymin=0.14,
xmin=1450,xmax=1630,
xticklabel style={/pgf/number format/1000 sep=},
    ytick distance=0.02,
    xtick distance=40,
]

\addplot[black,very thick] table[x=wav,y=attenuation] {Data/attenuation2.txt};

\node[anchor=west] (source) at (axis cs:1490-6,0.231){S};
\node[anchor=west] (source) at (axis cs:1543-6,0.231){C};
\node[anchor=west] (source) at (axis cs:1588-2,0.146){L};

 \draw[>=stealth, ->,line width=1.2pt] (axis cs: 1470,0.16-0.005)--(axis cs:1470,0.189-0.005);
\draw[>=stealth, ->,line width=1.2pt] (axis cs: 1499,0.16-0.005)--(axis cs:1499,0.170-0.005);
\draw[>=stealth, ->,line width=1.2pt] (axis cs: 1502,0.16-0.005)--(axis cs:1502,0.173-0.005);
\node[above] at (axis cs: 1485,0.14) {Raman pumps};
\begin{scope}
    \fill[red,opacity=0.1] ({rel axis cs:0.111111,0.0}) rectangle ({rel axis cs:0.38888888,1});
    \fill[green,opacity=0.1] ({rel axis cs:0.38888888,0}) 
    rectangle ({rel axis cs:0.6388888,1});
    \fill[blue,opacity=0.1] ({rel axis cs:0.63888888,0})
    rectangle ({rel axis cs:0.9166666,1});
\end{scope}

\end{axis}

 \begin{axis}[tiny,anchor=north east,at={(rel axis cs:1.1,0.98)}, anchor=north east,
 legend cell align=left,
legend style={font=\footnotesize, at={(rel axis cs:1,1)}, anchor=north east},
xmin=0,xmax=25,
yticklabel style={
        /pgf/number format/fixed,
        /pgf/number format/precision=5
},
xtick = {0,5,10,15,20,25},
ylabel={Raman gain [1/km/W]},
x label style={at={(axis description cs:0.5,-0.1)},anchor=north},
y label style={at={(axis description cs:-0.35,0.48)},anchor=north},
xlabel={Frequency separation [THz]},
ymin=0,ymax=0.2,
ytick = {0,0.05,0.1,0.15,0.2,0.25},
ytick distance=0.025,
grid=both,
]
  \begin{scope}[on background layer]
    \draw[fill=white] (-9,-0.07) rectangle (26,0.22);
  \end{scope}   
               \addplot[very thick] table[x=freq,y expr=\thisrowno{2}*(1)] {Data/Raman150.txt};
 \end{axis}

\end{tikzpicture}
\begin{tikzpicture}[baseline]
\begin{axis}[
legend cell align=left,
  legend columns=3,
legend style={font=\footnotesize, at={(rel axis cs:0,0)}, anchor=south west},
width=8.5cm, height = 5.5cm,
xlabel={Wavelength [nm]},
ylabel={Amplifier Gain [dB]},
grid=both,
unbounded coords=jump,
ymax=12,ymin=-2,
xmin=1520,xmax=1620,
xticklabel style={/pgf/number format/1000 sep=},
    ytick distance=2,
    xtick distance=20,
]

\begin{scope}[on background layer]
    \fill[green,opacity=0.1] ({rel axis cs:0.1,0.0}) rectangle ({rel axis cs:0.475,1});
    \fill[blue,opacity=0.1] ({rel axis cs:0.475,0}) 
    rectangle ({rel axis cs:0.97,1});
\end{scope}

\node[anchor=west] (source) at (axis cs:1540,11){C};
\node[anchor=west] (source) at (axis cs:1590,11){L};
    
\addlegendentry{Raman}
\addplot[Set1-A,very thick] table[x=wav,y=Raman] {Data/Gain.txt};

\addlegendentry{Lumped}
\addplot[Set1-B,very thick] table[x=wav,y=EDFA] {Data/Gain.txt};

\addlegendentry{Total}
\addplot[Set1-C,very thick] table[x=wav,y=total] {Data/Gain.txt};

\end{axis}
\end{tikzpicture}

%% file: Figures/SNR.tex
\begin{tikzpicture}[baseline]
\begin{axis}[
unbounded coords=jump,
title = (a) Hybrid Backward Raman Amplification,
legend style={font=\footnotesize, at={(rel axis cs:1,0)}, anchor=south east},
legend columns=3,
width=8.5cm, height = 5.5cm,
xlabel={Wavelength [nm]},
ylabel={$\text{SNR}$ [dB]},
grid=both,
ymax=22,ymin=12,
xmin=1520,xmax=1620,
xticklabel style={/pgf/number format/1000 sep=},
  ylabel near ticks,
    ytick distance=2,
    xtick distance=20,
]

\node[anchor=west] (source) at (axis cs:1545,21.2){C};
\node[anchor=west] (source) at (axis cs:1587,21.2){L};

\node[anchor=west] (source) at (axis cs:1540,14.3){};
\node (destination) at (axis cs:1540,17){};
\draw[->](source)--(destination);

\node[] at (axis cs: 1540,18.2) {99.22 \footnotesize Tbit/s}; 
\node[] at (axis cs: 1540,17.3) {20.4 \footnotesize dBm};

\begin{scope}[on background layer]
    \fill[green,opacity=0.1] ({rel axis cs:0.1,0.0}) rectangle ({rel axis cs:0.475,1});
    \fill[blue,opacity=0.1] ({rel axis cs:0.475,0}) 
    rectangle ({rel axis cs:0.97,1});
\end{scope}

\addlegendentry{NLI}
\addplot[Set1-B, very thick] table[x=wav,y=NLI] {Data/SNR_Hybrid.txt};
\addlegendentry{ASE}
\addplot[Set1-A, very thick] table[x=wav,y=ASE] {Data/SNR_Hybrid.txt};
\addlegendentry{Total}
\addplot[Set1-C, very thick] table[x=wav,y=total] {Data/SNR_Hybrid.txt};

\end{axis}
\end{tikzpicture}
\begin{tikzpicture}[baseline]
\begin{axis}[
unbounded coords=jump,
title = (b) Lumped Amplification,
legend style={font=\footnotesize, at={(rel axis cs:0,0.61)}, anchor=north west},
legend columns=1,
width=8.5cm, height = 5.5cm,
xlabel={Wavelength [nm]},
ylabel={$\text{SNR}$ [dB]},
grid=both,
ymax=22,ymin=12,
xmin=1520,xmax=1620,
xticklabel style={/pgf/number format/1000 sep=},
  ylabel near ticks,
    ytick distance=2,
    xtick distance=20,
]

\node[anchor=west] (source) at (axis cs:1545,21.2){C};
\node[anchor=west] (source) at (axis cs:1587,21.2){L};

\node[anchor=west] (source) at (axis cs:1582,13.2){};
\node (destination) at (axis cs:1584,16){};
\draw[->](source)--(destination);

\node[] at (axis cs: 1586,17) {88.55 \footnotesize Tbit/s};
\node[] at (axis cs: 1586,16.2) {23 \footnotesize dBm};

\begin{scope}[on background layer]
    \fill[green,opacity=0.1] ({rel axis cs:0.1,0.0}) rectangle ({rel axis cs:0.475,1});
    \fill[blue,opacity=0.1] ({rel axis cs:0.475,0}) 
    rectangle ({rel axis cs:0.97,1});
\end{scope}

\addlegendentry{NLI}
\addplot[Set1-B, very thick] table[x=wav,y=NLI] {Data/SNR_lumped.txt};
\addlegendentry{ASE}
\addplot[Set1-A, very thick] table[x=wav,y=ASE] {Data/SNR_lumped.txt};
\addlegendentry{Total}
\addplot[Set1-C, very thick] table[x=wav,y=total] {Data/SNR_lumped.txt};

\end{axis}
\end{tikzpicture}

%% file: main.bbl
\begin{thebibliography}{1}
\newcommand{\enquote}[1]{``#1''}

\bibitem{BugliaJOCN}
H.~Buglia~\textit{et al.}, {\protect\JournalTitle{Journal of Optical Communications and Networking}} \textbf{14}, B11--B21 (2022).

\bibitem{RamanJLT}
H.~Buglia~\textit{et al.}, {\protect\JournalTitle{Journal of Lightwave Technology, Early Access, DOI: 10.1109/JLT.2023.3315127}}  (2023).

\bibitem{ECOC_Raman}
H.~Buglia~\textit{et al.}, {\protect\JournalTitle{European Conference on Optical Communication (ECOC), P4}}  (2023).

\bibitem{bugliaOFC}
H.~Buglia~\textit{et al.}, {\protect\JournalTitle{Optical Fiber Conference (OFC), W2A.29}}  (2023).

\bibitem{torino}
M.~Ranjbar Zefreh~\textit{et al.}, {\protect\JournalTitle{Optical Fiber Conference (OFC), M5C.1}}  (2021).

\bibitem{Semrau_integral}
D.~Semrau~\textit{et al.}, {\protect\JournalTitle{J. Lightwave Technol.}} \textbf{36}, 3046--3055 (2018).

\bibitem{RamanEquations}
J.~Bromage, {\protect\JournalTitle{Journal of Lightwave Technology}} \textbf{22}, 79--93 (2004).

\bibitem{daniel_closed}
D.~Semrau~\textit{et al.}, {\protect\JournalTitle{Journal of Lightwave Technology}} \textbf{37}, 1924--1936 (2019).

\end{thebibliography}
